\documentclass[twocolumn, prb, aps, showpacs]{revtex4}        

\usepackage{graphicx}
\usepackage{wasysym}
\begin{document}
\title{Low-temperature specific heat and thermal conductivity of glycerol}
\author{C. Tal\'on}
\author{Q. W. Zou}
\author{M. A. Ramos}
\author{R. Villar}
\author{S. Vieira}
\affiliation{Laboratorio de Bajas Temperaturas, Departamento de F\'{\i}sica de la Materia Condensada, C-III \\
Instituto de Ciencia de Materiales ``Nicol\'as Cabrera'', Universidad Aut\'onoma de Madrid, Cantoblanco, E-28049 Madrid, Spain}
\date{\today}
\begin{abstract}
We have measured the thermal conductivity of glassy glycerol between 1.5~K and 100~K, as well as the specific heat of both glassy and crystalline phases of glycerol between 0.5~K and 25~K. We discuss both low-temperature properties of this typical molecular glass in terms of the soft-potential model. Our finding of an excellent agreement between its predictions and experimental data for these two independent measurements constitutes a robust proof of the capabilities of the soft-potential model to account for the low-temperature properties of glasses in a wide temperature range.
\end{abstract}  

\maketitle

It is well established \cite{ZP,stephens} that glasses or amorphous solids exhibit thermal properties 
very different from those of crystalline solids and, even more strikingly, very similar among themselves 
irrespective of the type of material, chemical bonding, etc. At temperatures $T < 1$~K, the specific 
heat $C_p $ of non-metallic glasses is significantly larger and the thermal conductivity $\kappa$ orders 
of magnitude lower than those found in their crystalline counterparts. $C_p$ depends approximately 
linearly ($C_p \propto T$) and $\kappa$ almost quadratically ($\kappa \propto T^{2}$) on temperature, in 
contrast to the cubic dependences observed in crystals for both properties, well understood in terms of 
Debye's theory. At $T > 1$~K, $C_p $ still deviates strongly from the expected 
$C_{\rm Debye} \propto T^3$ dependence, exhibiting a hump in $C_{p} / T^{3}$. In the same temperature 
range the thermal conductivity exhibits a universal {\em plateau}. 

In 1972, Phillips \cite{phil72} and Anderson, Halperin and Varma \cite{AHV} introduced independently 
the well-known  tunneling model (TM), whose fundamental postulate was the ubiquitous existence of small 
groups of atoms in amorphous solids which can tunnel between two configurations of very similar energy. 
This simple model of two-level systems or tunneling states successfully explained the low-temperature 
properties of amorphous solids \cite{phil_rev}, though only for $T < 1$~K. On the contrary, the also rich universal behavior of glasses above 1~K (the hump in $C_{p} / T^{3}$ and the plateau in the thermal conductivity, or the remarkable feature in the vibrational density of states $g (\nu ) / \nu^2$ at low frequencies known as the {\em boson peak}) 
still remains a matter of debate. One of the best accepted approaches to understand all the general 
behavior of glasses in the whole range of low-energy excitations is the phenomenological soft-potential 
model (SPM), which can be regarded as an extension of the TM. The SPM \cite{kki,ikp} postulates the 
coexistence of extended lattice vibrations (sound waves) with quasilocalized low-frequency ({\em soft}) 
modes. In this model, the potential of these soft modes has a uniform stabilizing fourth-order term $W$. 
In addition, each mode has its individual first-order asymmetry $D_1$ and second-order restoring force 
terms $D_2$, which can be either positive or negative. Similarly to the TM, a random distribution of 
potentials is assumed: $P (D_1 , \, D_2) = P_s $. The SPM has been developed 
\cite{kki,ikp,bggprs,grbb,gurevich,rb} and reviewed \cite{parshin,esqui} in earlier papers where the 
interested reader is referred to.

Glycerol [C$_3$H$_5$(OH)$_3$] is probably the most widely studied \cite{wuttke} glass-forming liquid. 
Its high viscosity at a melting point around room temperature ($T_{m} = 291 $~K) provides 
experimentalists with a very convenient temperature range where the supercooled liquid can be studied. 
Below the glass transition at $T_{g} \simeq 185 $~K, the frozen-in liquid becomes a glass with a 
relatively weak, hydrogen-bonded network structure. Despite its good glass-forming ability, glycerol 
can also be obtained in an orthorhombic crystalline state \cite{koning}, with four C$_3$H$_5$(OH)$_3$ 
molecules per unit cell, building up a structure of infinite hydrogen-bonded chains \cite{FJB}. Several 
measurements of the specific heat of this well-known glass have been indeed reported 
\cite{ahlberg,craig,leadbetter,calemczuk}, though not reaching temperatures below 1.5~K in any case. A broad maximum in $C_{p} / T^{3}$ was clearly observed \cite{leadbetter,calemczuk} around 8.5~K, but the expected 
existence of tunneling states could not be determined, since it requires temperatures typically 
below 1~K. Furthermore, Calemczuk {\em et al.} \cite{calemczuk} also measured the specific heat of 
the crystalline state of glycerol, but only down to 5~K, hence not reaching temperatures low enough 
as to assess its Debye coefficient. On the other hand, there are no published thermal-conductivity 
data at low temperatures neither for glycerol nor for any other similar molecular glass. This is very 
probably due to the experimental difficulties of adapting the thermal-conductivity technique 
to a sample which is liquid at ambient temperature and must be thermally controlled {\em in situ} to 
freeze it into the glass state, maintaining at the same time an appropriate geometry and a moderate 
heat flow for the thermal conductivity to be correctly measured. To our knowledge, only a few 
orientationally-disordered (``glassy'') {\em crystals} from other molecular liquids have been 
measured \cite{bonjour,vieira}.  

In this work, we report thermal-conductivity data of glassy glycerol measured from 1.5~K to around 
100~K, as well as specific-heat measurements of both glassy and crystalline glycerol between 0.5~K 
and 25~K. In addition, we make concurrent use of both thermal properties measured for the glassy 
state in order to test the validity of the soft-potential model in a typical molecular glass.  

\begin{figure}[t]
\includegraphics[width=7.5cm,clip]{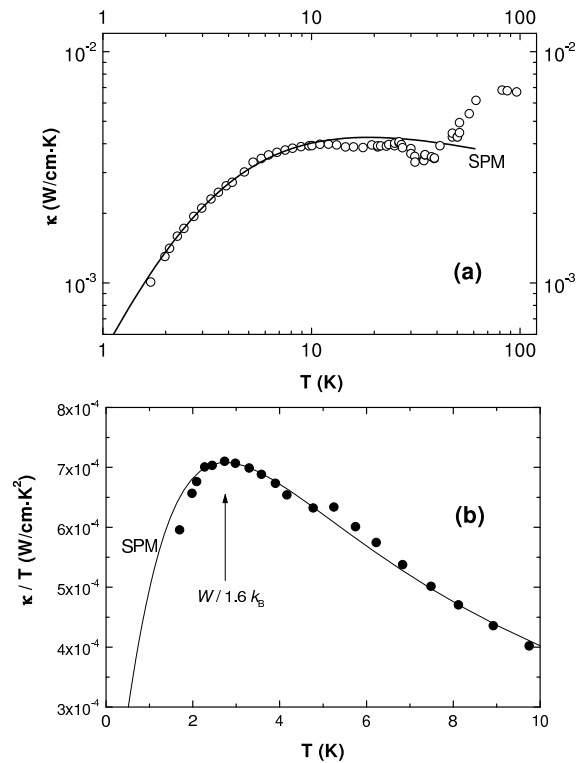}
\caption{Low-temperature thermal conductivity of glassy glycerol. (a): Thermal conductivity as a 
function of temperature in a log-log plot. Solid line is a fit to the soft-potential model (SPM). 
(b): Thermal conductivity divided by temperature below 10 K. The maximum marks the crossover from 
tunneling states to quasilocalized vibrations (soft modes) as dominant scattering centers for the 
``phonons'' and allows a direct determination of the SPM parameter $W$.}
\label{conductividad}
\end{figure}

In Fig.~\ref{conductividad}(a), we show our thermal-conductivity data of glassy glycerol obtained by 
cooling {\em in situ} below $T_{g}$ at a rate around -1 K/min liquid glycerol (Merck, anhydrous, used 
without futher purification), placed inside a very thin-walled (0.2 mm) nylon tube. Standard 
steady-state techniques were employed. The thermal conductivity of the empty nylon tube was 
independently measured and subtracted. As can be seen, glycerol exhibits the thermal-conductivity 
behavior typical of glasses, with a plateau around 10~K quantitatively very similar to that of strong, 
network glasses as As$_2$S$_3$ and GeO$_2$ \cite{stephens}, i.e., a relatively high thermal 
conductivity among glasses. Therefore, glycerol could be a useful heat exchange medium at low 
temperatures when used as a glassy matrix.

Let us first analyze the thermal conductivity of glycerol within the soft-potential model. According 
to the SPM
\cite{bggprs}, the inverse mean-free path of the phonons carrying out the heat is simply the sum of 
three contributions: the resonant scattering of the sound waves by either tunneling states or by 
{\em soft} vibrational modes, and the scattering by classical relaxational processes in the same 
asymmetric double-well potentials responsible of the tunneling states. From these premises, and 
inserting the corresponding SPM expressions for these three inverse mean-free paths, the thermal 
conductivity $\kappa$ was found to be \cite{rb,esqui}
\begin{equation}
\label{conduct}
\kappa =\frac{2k_{\rm B}}{3\pi \overline{C}}\left( \frac {W}{h}\right) ^2 F(z),
\end{equation}
where
\begin{equation}
\label{fu}
F(z)=\int_0^\infty dx\frac{x^3{\rm e}^{-x}}{(1-{\rm e}^{-x})^2} 
\frac{z^2}{1.1 \tanh(x/2)+0.7z^{3/4}+x^3z^3/8} 
\end{equation}

$\overline{C}$ is the usual (averaged over longitudinal and transverse acoustic modes) dimensionless 
constant of the TM, directly related to the universal plateau in the internal friction 
$Q^{-1} = (\pi / 2) C$, and $z = k_{\rm B}T / W$, where the energy $W$ is the aforementioned parameter of the SPM, which marks the crossover from the tunneling-states region at the lowest 
temperatures to the soft-modes region above it. Indeed, $W$ can be determined 
\cite{rb,esqui} from the position of the maximum $T_{max}$ in a $\kappa/T$ versus $T$ plot:
\begin{equation}
W  \simeq \, 1.6 \, k_{\rm B} T_{max}   .
\end{equation}

$T_{max}$ therefore separates the low-temperature range where resonant scattering by tunneling 
states dominates ($\kappa \propto T^2$) from higher temperatures 
($\kappa \approx {\rm const}$) where the soft modes are the main scatterers of acoustic phonons. 

Fig.~\ref{conductividad} shows that SPM eqs. (\ref{conduct}-\ref{fu}) provide a good fit to 
thermal-conductivity data of glycerol from the lowest temperatures up to the end of the plateau 
around 30~K. The two parameters, $W$ and $\overline{C}$  are easily obtained \cite{rb,esqui}
from the $\kappa/T$ plot [Fig.~\ref{conductividad}(b)]: the maximum position gives $W = 4.3$~K and its height 
$\overline{C} = 1.9 \times 10^{-4}$. The latter constitutes indeed a prediction for the 
low-temperature internal friction of glycerol, never measured to our knowledge: its plateau value 
should be around $Q^{-1} \approx 3 \times 10^{-4}$, up to \cite{rb,esqui} 
$ 1.2 W / k_{\rm B} \, \simeq \, 5$~K, where the rise due to thermal relaxation should occur.

\begin{figure}[ht]
\includegraphics[width=5cm,angle=270,clip]{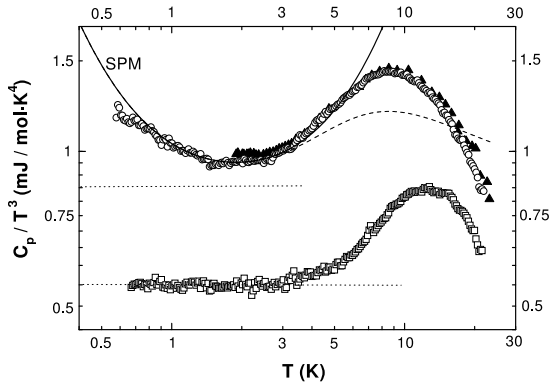}
\caption{Low-temperature specific heat $C_{p} / T^{3}$ of glass $(\bigcirc)$ and crystal $(\Box)$ 
phases of glycerol for  the whole measured temperature range in a log-log scale. Solid triangles are 
published data from Leadbetter and Wycherley 
{\protect\cite{leadbetter}}. Dotted lines show the correspondent Debye contributions to the specific 
heat either measured calorimetrically (crystal) or estimated from sound velocities (glass). Solid line 
shows the curve calculated with the soft-potential model (SPM), taking the parameter $W$ from the 
thermal conductivity data. Dashed line shows the same SPM calculation, but with a gaussian cutoff of 
the asymmetry for soft modes at high energies (see text).}
\label{Cp-1}
\end{figure}

\begin{figure}[ht]
\includegraphics[width=5cm,angle=270,clip]{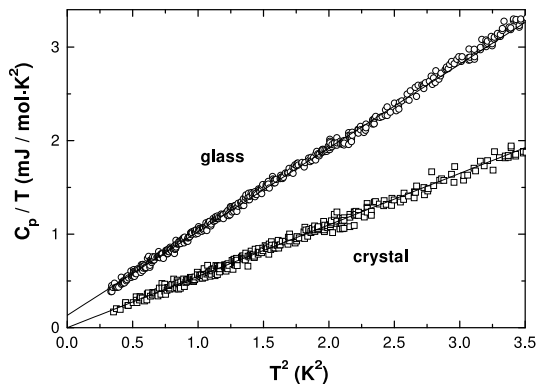}
\caption{Low-temperature specific heat of glass and crystal phases of glycerol in a $C_{p} / T$ vs 
$T^{2}$ plot. Symbols are as in Fig.~\ref{Cp-1}. Solid lines are least-squares linear fits.}
\label{Cp-2}
\end{figure}

Our specific-heat measurements of glassy and crystalline states of glycerol are presented in 
Fig.~\ref{Cp-1} (in a log-log plot showing the whole low-temperature range) and in Fig.~\ref{Cp-2} 
(in a $C_{p} / T$ vs $T^{2}$ plot at the lowest temperatures to address the tunneling-states range 
for the glass and the Debye limit for the crystal). 
The heat capacity was measured in a $^3$He-cryostat, employing a low-temperature quasi-adabiatic 
calorimetric cell, similar to one previously used in a $^4$He-cryostat \cite{EtOH,POH}. The liquid glycerol is placed in a vacuum-tight, thin-walled copper can, with a fine mesh of copper fitted inside to facilitate thermal equilibrium. A thin gold wire was used as heat switch to cool the experimental cell. The subtracted addenda contibution to the total measured heat capacity at 4.2~K (1~K) was about 15~\% (22~\%) for the glass and about 30~\% (50~\%) for the crystal. The glass state was obtained by simply cooling the liquid from room temperature down to liquid-helium temperatures. After having measured its heat capacity at low temperature, glassy glycerol was slowly heated above $T_{g}$ (calorimetrically observed to take place at 185~K) until it crystallized around 250--260~K. Once this first-order, exothermic transition was completed, the sample was cooled again and the heat capacity of the crystal measured.

In Fig.~\ref{Cp-1}, published data for the glass between 1.9 and 25 K from Leadbetter and Wycherley 
\cite{leadbetter} can be seen to show a very good agreement with our data. There is also a good agreement with data for the crystal above 5~K \cite{calemczuk}. The low-temperature cubic 
Debye contributions are also indicated in that figure: for the crystal, it has been obtained from a 
least-squares linear fit below 1.5~K in Fig.~\ref{Cp-2}, giving  a Debye temperature 
$\Theta_{\rm D} = 367 \,$K; for the glass, it can be estimated from zero-temperature extrapolations 
of elastic data \cite{schulz,litovitz,jeong} to be $\Theta_{\rm D} = 317 \,$K. In contrast to the 
crystal, where $C_{p}$ holds the expected Debye behavior for $T \leq \Theta_{\rm D} / 50$, the 
glass exhibits the typical excess over the Debye contribution in the whole low-temperature range. 
Within the SPM, these additional excitations are understood as soft quasiharmonic vibrations 
in single-well potentials responsible for the ``boson peak'' in $C_{p} / T^{3}$, together with tunneling 
states arising from related double-well potentials responsible for the quasilinear contribution. The latter dominates below the minimum $T_{\rm min}$ in $C_{p} / T^{3}$ given by $W  \approx \, 1.8 - 2 \, k_{\rm B} T_{\rm min} $ \cite{parshin,esqui}, and is more clearly observed in Fig.~\ref{Cp-2}, where a linear coefficient $\gamma = 0.13 \,$mJ/mol$\cdot$K$^2$ can be obtained (i.e., $\gamma = 1.4 \, \mu $J/g$\cdot$K$^2$ that is very similar to those values found in typical network glasses \cite{ZP,stephens,phil_rev}).  

Finally, we wish to compare the specific-heat data of glassy glycerol with the behavior predicted by 
the SPM, once we have previously obtained its basic parameter $W = 4.3$~K from thermal-conductivity 
data. As said above, the SPM postulates the coexistence of extended sound waves with quasilocalized modes, either tunneling states in double-well potentials or quasiharmonic vibrations in single-well potentials, with a gradual crossover between them characterized by the energy $W$. Therefore, we can write the specific heat for a glass as 
\begin{equation}
\label{suma}
C_p  \, = \, C_{\rm Debye} \, + C_{\rm TLS} \, + C_{\rm sm}  ,
\end{equation}
where $C_{\rm sm}$ is the contribution of soft modes (see below), and the contribution of the tunneling 
states $C_{\rm TLS}$ is determined, as in the TM, by means of the well-known expression for two-level 
systems \cite{phil_rev}, using the correspondent density of 
tunneling states in terms of the SPM (eq. (4.7) in Ref. \cite{parshin} or eq. (9.30) in Ref. 
\cite{esqui}). 

In the standard SPM, the density of quasiharmonic soft vibrations increases continuously with frequency as
\begin{equation}
\label{dos}
g_{\rm sm} (h \nu ) = \frac {1}{8} \frac {P_s}{W} \left( \frac {h\nu}{W} \right)^4 ,
\end{equation}
where $P_s$ is the distribution constant of soft potentials.
In the harmonic approximation, this leads to a specific-heat contribution from soft modes given by 
\cite{esqui}
\begin{equation}
\label{cp-sm}
C_{\rm sm} = \frac{2 \pi^6}{21} P_s k_{\rm B} \left(\frac{k_{\rm B}T}{W}\right)^{5} .
\end{equation}

The solid line in Fig.~\ref{Cp-1} shows the result of eq. (\ref{suma}), where $W = 4.3$~K was taken 
from the thermal-conductivity measurement and $P_s = 1.6 \times 10^{19}$ mol$^{-1}$ was simply 
determined to scale with experimental data. 

Obviously, the {\em real} distribution of soft modes cannot grow with frequency 
$g_{\rm sm} (\nu ) \propto \nu^4$ unlimitedly.
Gil {\em et al.} \cite{grbb} proposed a gaussian distribution in the asymmetry of the soft potentials 
(hence multiplying eq. (\ref{dos}) by an integral factor, see eq. (9.40) in Ref. \cite{esqui}) based 
in a thermal strain {\em ansatz}, which without any further fitting parameter allowed them to account 
for the specific heat, thermal conductivity and vibrational density of states $g (\nu ) / \nu^2$ in 
the whole relevant range, including the ``boson peak''. Alternatively, Gurevich {\em et al.} 
\cite{gurevich} argued that the simple picture of independent quasilocalized harmonic vibrations 
coexisting with sound waves should fail at the Ioffe-Regel limit, not far above the boson peak. The 
interaction between soft modes would lead to a reconstruction of the vibrational density of states at 
higher frequencies, where delocalized soft vibrations with $g_{\rm sm} (\nu ) \propto \nu$ should 
dominate, therefore also explaining the boson-peak feature.
Furthermore, hybridization of acoustic phonons with quasilocalized modes has been proposed to set in 
around the boson peak \cite{elliott}. For the sake of completeness, we also show in Fig.~\ref{Cp-1} by 
a dashed line the result of the total SPM prediction for the specific heat if the above-mentioned 
correction \cite{grbb,esqui} of eq. (\ref{cp-sm}) is used. As can be seen, the position of the maximum in 
$C_{p} / T^{3}$ (which only depends on $W$ and the glass transition temperature $T_{\rm g}$, hence being determined without any free parameter) is very well predicted, though its height is not so well accounted for.
However, these quantitative agreements or disagreements of the SPM around or above the boson peak are 
perhaps not very relevant, since the low-energy limits of both independent quasilocalized modes and Debye 
acoustic phonons (note that even in the crystal, $C_{p}$ starts to deviate from the cubic limit above 
5 K) should begin to fail there, for the reasons mentioned above. Nevertheless, one may conclude that 
the maximum in $C_{p} / T^{3}$ of glasses (i.e., the boson peak) is just the fingerprint of the end for 
the low-energy distribution of independent soft modes, which govern low-temperature properties and 
low-frequency dynamics of glasses.

In any case, it is noteworthy that with only a constant factor $P_s$, used as free parameter to fit 
the height of the $C_{p} (T) $ curve ($W$ was independently determined from thermal-conductivity data), 
the SPM is able to account consistently for the specific heat of this glass in the low-temperature range, 
from the tunneling-states region below 1~K up to the broad peak in $C_{p} / T^{3}$, including the 
crossover region around the minimum in $C_{p} / T^{3}$. 

In summary, we have concurrently measured two low-temperature thermal properties of a paradigmatic molecular glass, glycerol, as well as the specific heat of the crystalline phase between 0.5~K and 25~K, hence being able to determine its Debye temperature. At lower temperatures, both properties for the glass exhibit a typical behavior indicative of the existence of tunneling states. Moreover, we have used these data as a new test of the SPM, which has been shown to successfully explain the specific heat and the thermal conductivity in a wide temperature range, also for a molecular glass such as glycerol.
\acknowledgments
This work was supported by MCyT (Spain) within project BFM2000-0035-C02-01.

\end{document}